\documentclass[a4paper,english,aps, numbers, sort&compress, twocolumn]{revtex4-1}
\usepackage[T1]{fontenc}
\usepackage[latin9]{inputenc}
\usepackage{color}
\usepackage{amsmath}
\usepackage{amssymb}
\usepackage{graphicx}

\makeatletter

\@ifundefined{textcolor}{}
{%
 \definecolor{BLACK}{gray}{0}
 \definecolor{WHITE}{gray}{1}
 \definecolor{RED}{rgb}{1,0,0}
 \definecolor{GREEN}{rgb}{0,1,0}
 \definecolor{BLUE}{rgb}{0,0,1}
 \definecolor{CYAN}{cmyk}{1,0,0,0}
 \definecolor{MAGENTA}{cmyk}{0,1,0,0}
 \definecolor{YELLOW}{cmyk}{0,0,1,0}
}
\numberwithin{equation}{section}

\makeatother

\usepackage{babel}
\begin{document}

\title{Emergence and combinatorial accumulation of jittering regimes in
spiking oscillators with delayed feedback}

\author{{\normalsize{}Vladimir~Klinshov$^{1}$, Leonhard~L�cken$^{2}$,
Dmitry Shchapin$^{1}$, Vladimir~Nekorkin$^{1,3}$, and Serhiy~Yanchuk$^{2}$}}

\affiliation{$^{1}$Institute of Applied Physics of the Russian Academy of Sciences,
46 Ul'yanov Street, 603950, Nizhny Novgorod , Russia}

\affiliation{$^{2}$Weierstrass Institute for Applied Analysis and Stochastics,
Mohrenstrasse 39, 10117, Berlin, Germany}

\affiliation{$^{3}$University of Nizhny Novgorod, 23 Prospekt Gagarina, 603950,
Nizhny Novgorod, Russia}


\date{\today}
\begin{abstract}
Interaction via pulses is common in many natural systems, especially
neuronal. In this article we study one of the simplest possible systems
with pulse interaction: a phase oscillator with delayed pulsatile
feedback. When the oscillator reaches a specific state, it emits a
pulse, which returns after propagating through a delay line. The impact
of an incoming pulse is described by the oscillator's phase reset
curve (PRC). In such a system we discover an unexpected phenomenon:
for a sufficiently steep slope of the PRC, a periodic regular spiking
solution bifurcates with several multipliers crossing the unit circle
at the same parameter value. The number of such critical multipliers
increases linearly with the delay and thus may be arbitrary large.
This bifurcation is accompanied by the emergence of numerous ``jittering''
regimes with non-equal interspike intervals (ISIs).  Each of these
regimes corresponds to a periodic solution of the system with a period
roughly proportional to the delay. The number of different ``jittering''
solutions emerging at the bifurcation point increases exponentially
with the delay. We describe the combinatorial mechanism that underlies
the emergence of such a variety of solutions. In particular, we show
how a periodic solution exhibiting several distinct ISIs can imply
the existence of multiple other solutions obtained by rearranging
of these ISIs. We show that the theoretical results for phase oscillators
accurately predict the behavior of an experimentally implemented electronic
oscillator with pulsatile feedback.
\end{abstract}
\maketitle

\section{Introduction}

Oscillating systems subject to pulsed inputs or interactions were
studied in many different areas, such as dynamics of spiking neurons
\citep{kunysz1997bursting}, communication of fireflies by short light
pulses \citep{Mirollo1990,Winfree2001}, impacting mechanical oscillators
\citep{Brzeski2015}, electronic oscillators \citep{Lopera2006,Rosin2013,Klinshov2014},
optical systems \citep{Colet1994,Boyd2009,Otto2012}, stimulation
of cardiac \citep{Michaels1984,Dokos1996,Tsalikakis2007}, respiratory
\citep{Lewis1987,Lewis1992} or circadian \citep{Minors1991} rhythms.
Simple but powerful models to describe such systems are phase oscillators
with pulsatile coupling, which are especially popular in neuroscience
\citep{Mirollo1990,Abbott1993,Tsodyks1993,Ernst1995,Bottani1995,Vreeswijk1996,Bressloff1997,Ernst1998,Goel2002,Rossoni2005,Zillmer2007,Achuthan2009,Canavier2010,LaMar2010,Lucken2012a,Luecken2013}.
Apart of the simplicity of phase models in comparison to conductance-based
models \citep{Izhikevich2005}, they possess two main features: the
possibility of individual neurons to produce periodic output, and
the fact that interaction between neurons is mediated by the brief
action potentials or spikes, which have a temporal duration much smaller
than the interspike intervals (ISIs). Moreover, the effect of a spike
depends on the dynamical state at which the neuron is located at the
time of the spike arrival. In phase models such an effect is incorporated
with the help of phase reset curves (PRC) \citep{Kuramoto1984,Ermentrout1996,Tass1999,Winfree2001,Canavier2010}.
In its representation as a phase oscillator, each type of neuron possesses
a characteristic PRC corresponding to a particular stimulus. Hence,
parameter changes in the neuron or the stimulus are reflected by changes
in the shape of the PRC. PRCs can be computed for any oscillatory
system and stimulus including neuronal models such as Hodgkin-Huxley,
FitzHugh-Nagumo, and others \citep{Ermentrout1996,Novicenko2011}.
In this way, pulse-coupled systems can be considered either as stand-alone
models, or as approximations of coupled oscillatory conductance-based
systems. Among the advantages of such models is their simple numerical
implementation, their lower dimension in comparison with conductance-based
models, as well as the possibility to adjust the PRC numerically and
measure it experimentally \citep{Glass1984,Anumonwo1991,Galan2005,Ermentrout2011,Foss2000}.
The phase description can serve as an appropriate approximation, if
the responses of the system to inputs introduce transient deviations
from the oscillatory state, which are shorter than the ISIs.

In this paper we consider a single oscillator with pulsatile delayed
feedback extending the results reported in \citep{Klinshov2015a}.
The study on this basic  and very common ``motif'' \citep{y1995histology}
is important for understanding the behavior of larger delay-coupled
networks \citep{Milo2002}. For instance, a loop consisting of one
excitatory and one inhibitory neuron with delayed connection shows
similar behavior to a neuron with delayed self-feedback \citep{Ma2007,Ma2010,Hashemi2012},
and also the behavior of rings of several neurons are, in some cases,
related to the behavior of a single neuron with delayed feedback \citep{VanderSande2008,Kantner2013,Yanchuk2011}.
In fact, a larger neuronal feedback delay might firstly arise due
to a chained propagation of action potentials along a ring of neurons.
Another motivation to study a single oscillator with delayed feedback
arises from the study of ensembles of delay coupled oscillators where
the synchronous regime and some modes of its instability can be described
by the behavior of a single oscillator  \citep{Flunkert2010,timme2008simplest}. 

Although we approach the problem of an oscillator with pulsed feedback
from a general perspective, not bound to a specific area of application,
neurons with delayed feedback are a prototypical example for such
systems. There exist some studies on this subject \citep{Foss1996,Foss2000,Ma2007,Ma2010,Hashemi2012},
all of which have a common baseline: delay leads to immense multistability.
At first glance this result is not too surprising since multistability
generically arises in delay differential equations due to a well-known
mechanism called ``periodic solution reappearance'' \citep{Yanchuk2009}.
Interestingly, in the case of a pulsatile feedback, this mechanism
does not always explain the large variety of observed solutions. Another
property, which is connected to some combinatorial relations between
the ISIs is responsible for the extreme multistability which arises
for larger values of the feedback delay. For example, Ma and Wu \citep{Ma2007,Ma2010}
found for several neuron models that a large number of periodic solutions
coexists. Remarkably, these solutions exhibit only a few distinct
ISIs; moreover, different solutions can be transformed into another
by permuting the order of the ISIs. Our results shed some light on
these earlier works from another perspective.

We show that an oscillatory system with delayed pulsatile feedback
may generically exhibit a very surprising phenomenon, which manifests
itself as follows: under some conditions a periodical regime of regular
spiking destabilizes in a degenerate manner such that several multipliers
(Lyapunov exponents) become critical at once. The number of the critical
multipliers is proportional to the feedback delay and can be arbitrary
large. Thus, the dimension of the unstable manifold of the regular
spiking solution changes abruptly from zero to an arbitrary large
value, which we call the ``dimension explosion'' phenomenon. As
a result of such a bifurcation, we show that there appear multiple
coexistent  periodic solutions with larger periods, characterized
by distinct ISIs. We call such regimes with non-equal ISIs ``jittering''
regimes, by analogy with well-known timing jitter in optical and electronic
systems \cite{Otto2012,derickson1991comparison,weigandt1994analysis}.
Previously, the solutions with non-equal ISIs have been reported in
various oscillator models subject to delayed feedback \cite{Ma2007,Ma2010,kunysz1997bursting,Lewis1987,Lewis1992,Foss1997}.
In our model, remarkably, the numerous jittering regimes emerge at
once at the bifurcation point which leads us to adopt the name ``multi-jitter''
bifurcation. We prove that the number of the emergent jittering solutions
grows exponentially with the delay.

The structure of the paper is as follows: In section~\ref{sec:Model}
we introduce the model for a phase oscillator with delayed pulsatile
feedback, introduce some definitions and assumptions, and give the
system's reformulation as discrete return map of ISI-sequences. In
section~\ref{sec:Regular-spiking} we study the regime of regular
spiking (RS) where all ISIs are identical. We obtain an explicit parametric
expression for all branches of RS solutions, determine their stability,
and give conditions for the multi-jitter bifurcation. It turns out
that the crucial quantity involved is the steepness of the PRC. In
the following Section~\ref{sec:Numerical} we undertake a numerical
exploration of the system's behavior beyond the multi-jitter bifurcation
and report a huge multistability of periodic solutions and chaotic
attractors. The most dominant type of the observed solutions exhibits
a periodic repetition of only two different ISIs in seemingly arbitrary
order. We explain this phenomenon in section~\ref{sec:Solution_explosion}
where we prove that the phenomenon of dimension explosion is indeed
accompanied by an explosion of the number of coexisting solutions.
In Section~\ref{sec:Experimental-observation-of} we present an experimental
realization, where the effects predicted by the phase model can be
observed in an electronic oscillator. We conclude with a discussion
in section~\ref{sec:Discussion}.

\section{Model\label{sec:Model}}

We consider an oscillator with delayed pulsatile feedback of the form
\citep{Ermentrout1991,Goel2002,Canavier2010,Achuthan2009,Lucken2012a,Luecken2013,Klinshov2015a}:
\begin{eqnarray}
\frac{d\varphi}{dt} & = & 1+Z(\varphi)\sum\limits _{t_{j}}\delta(t-t_{j}-\tau).\label{eq:1}
\end{eqnarray}
The oscillator is described by its phase $\varphi$, which changes
on the circle $[0,1]$ with $\varphi=0$ and $\varphi=1$ identified.
In the case without delayed feedback, the phase grows uniformly with
normalized frequency $\omega=1$. When the phase reaches unity, the
oscillator is assumed to emit a pulse. The instants when the pulses
are emitted are denoted by $t_{j}$, $j=1,2,\dots$. The emitted pulses
propagate along the feedback line and affect the oscillator after
the delay $\tau$ at the time instants $t_{j}^{\ast}=t_{j}+\tau$.
When a pulse is received, the phase of the oscillator undergoes an
instantaneous, discontinuous shift, and changes its value to the new
value
\[
\varphi(t_{j}^{\ast}+0)=\varphi+Z\left(\varphi\right),
\]
where $\varphi=\varphi(t_{j}^{\ast}-0)$ and the function $Z(\varphi)$
is the PRC \citep{Canavier2010,tsubo2007layer,Schultheiss2010}. In
the following, we suppose that $Z(\varphi)$ is continuous and differentiable
for all $\varphi$. Further, we assume that
\begin{equation}
Z(0)=Z(1)=0,\label{eq:Z1}
\end{equation}
which means that the oscillator does not respond to perturbations
during its own spike. In some cases, such an assumption is reasonable,
especially for the modeling of neuronal dynamics involving a fast
change in the systems state around $\varphi=0$, see additional discussion
in \citep{Lucken2012a,Luecken2013}. We also impose that
\begin{equation}
\varphi+Z(\varphi)\in[0,1]\label{eq:Z2}
\end{equation}
for all $\varphi\in[0,1]$.

\begin{figure}
\begin{centering}
\includegraphics{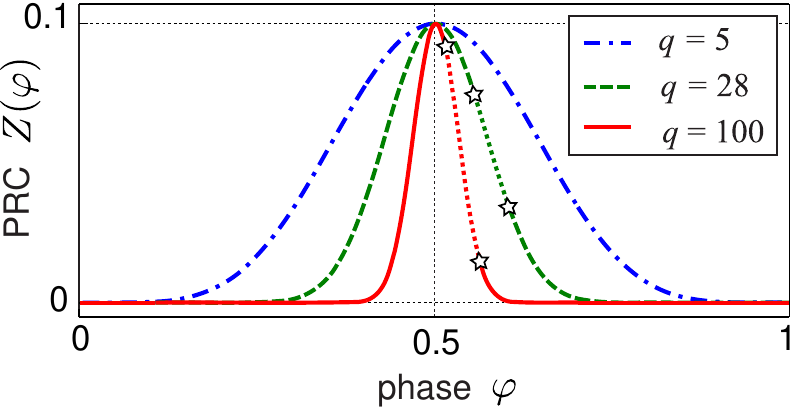} 
\par\end{centering}

\protect\caption{\label{fig:PRC}(Color online) PRC function (\ref{eq:PRC}) for $\kappa=0.1$
and different values of the parameter $q$. Stars indicate points,
where $Z^{\prime}(\varphi)=-1$ corresponding to possible multi-jitter
bifurcations {[}see Sec.~\ref{sec:Linear-stability}{]}. }
\end{figure}

For numerical illustrations we consider the PRC 
\begin{equation}
Z(\varphi)=\kappa\times\left(\sin\left(\pi\varphi\right)\right)^{q},\label{eq:PRC}
\end{equation}
which is shown in Fig.~\ref{fig:PRC}. Here $\kappa>0$ is the feedback
strength, and $q>1$ is a parameter that controls the steepness of
the PRC, which is the crucial quantity for the dynamical phenomena
reported in this paper.

 We define the steepness as the maximal \emph{downward} slope of $Z$:
\begin{equation}
s:=\left|\inf_{\varphi\in(0,1)}Z^{\prime}\left(\varphi\right)\right|.\label{eq:steepness-defenition}
\end{equation}
We emphasize, that the amplitude of the PRC and many features of
its shape (e.g., whether it is of type I or II, unimodal, etc.) are
not crucial for our results and the reported phenomena can be observed
for any PRC possessing a sufficiently high steepness. For the function
(\ref{eq:PRC}) the steepness can be estimated for large $q$ as
\begin{equation}
s\approx\kappa\times\pi\sqrt{q/e},\label{eq:steepness-formula}
\end{equation}
where $e$ is Euler's number {[}see Appendix~\ref{sec:Appendix-Steepness}{]}.

As for any system with time delay, some amount of information about
its past is required for system (\ref{eq:1}) in order to determine
its future evolution. However, because of the pulsatile nature of
the feedback, the only information needed is the time-moments of the
pulses $t_{j}$ (cf. \citep{Foss1997a,Losson1993}). In \citep{Klinshov2013}
it was proven that a system with pulsatile delayed coupling can be
reduced to a finite-dimensional map under quite general conditions.
For the oscillator (\ref{eq:1}) such a map can be obtained for the
ISIs $T_{j}=t_{j}-t_{j-1}$: 
\begin{equation}
T_{j+1}=F\left(T_{j},T_{j-1},...,T_{j-P+1}\right).\label{eq:map-definition}
\end{equation}
The map (\ref{eq:map-definition}) determines the next ISI based on
the $P$ preceding ISIs. Equation (\ref{eq:map-definition}) can equivalently
be written as the $P$-dimensional map\textcolor{red}{{} }
\[
\left(T_{1},...,T_{P}\right)\mapsto\left(T_{2},...,T_{P},F(T_{1},...,T_{P})\right).
\]
According to\textcolor{red}{{} }\citep{Klinshov2013}, the map dimension
is bounded as follows:
\[
P\leq1+\frac{\tau}{1-2Z_{\max}},
\]
where $Z_{\max}=\max_{0\le\varphi\le1}(Z(\varphi)).$ For the PRC
(\ref{eq:PRC}), $Z_{\max}=\kappa$.

\section{Regular spiking\label{sec:Regular-spiking}}

The map (\ref{eq:map-definition}) can have different and complicated
forms depending on the relation between the delay time and the spiking
frequency \citep{Klinshov2013}. Therefore, we will not write down
the general form of $F$ here. Instead, we derive the form of the
map for the case when there is only one feedback spike arriving at
the oscillator during each ISI. This form is adequate in many cases,
and in particular for RS regimes, which are fixed points of (\ref{eq:map-definition}),
and are characterized by a repeated spiking with a constant ISI length
$T_{j}\equiv T$ for all $j$. However, if the possibility of several
incoming pulses per ISI is taken into account, additional interesting
regimes can be observed \citep{Foss1997}.
\begin{figure}
\begin{centering}
\includegraphics{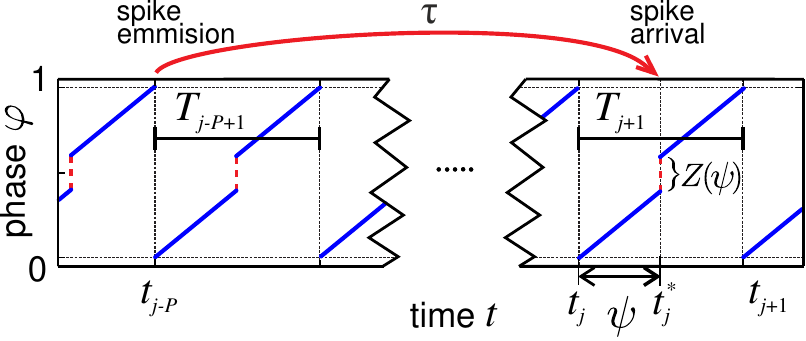}
\par\end{centering}

\protect\caption{\label{fig:map}(Color online) Spiking dynamics schematically. The
spike emitted at $t_{j-P}$ affects the dynamics after time delay
$\tau$ at $t_{j}^{*}=t_{j-P}+\tau$.}
\end{figure}

Under the assumption that exactly one spike arrives per ISI, let us
consider the oscillator dynamics on the time interval $[t_{j},t_{j+1}]$.
First, note that there exists a unique number $P$ of ISIs between
$t_{j}$ and the emission time $t_{j-P}$ of the spike arriving at
$t_{j}^{\ast}\in[t_{j},t_{j+1}]$, cf. Fig.~\ref{fig:map}. Accordingly,
we have
\[
t_{j}^{*}=t_{j-P}+\tau=t_{j}+\tau-\sum_{k=j-P+1}^{j}T_{k}.
\]
Further, the oscillator's phase $\varphi$ grows from zero to one
within the interval $[t_{j},t_{j+1}]$, and, except from the instant
$t_{j}^{\ast}$ when a feedback pulse arrives and perturbs the phase,
it grows linearly with time. The phase value $\psi\in\left[0,1\right]$
at the moment of the pulse arrival equals
\begin{equation}
\psi=t_{j}^{\ast}-t_{j}=\tau-\sum_{k=j-P+1}^{j}T_{k}.\label{eq:spike-arrival-offset-RS-1}
\end{equation}
Hence the corresponding shift of the perturbed phase equals $Z(\psi)$
and the oscillator's phase immediately after the pulse impact is $\varphi(t_{j}^{\ast}+0)=\psi+Z(\psi)$.
Finally, the ISI 
\[
T_{j+1}=t_{j+1}-t_{j}=\psi+t_{j+1}-t_{j}^{\ast}
\]
is determined from the condition $\varphi(t_{j+1})=1$:
\begin{align*}
1=\varphi(t_{j+1}) & =\psi+Z(\psi)+t_{j+1}-t_{j}^{\ast}\\
 & =T_{j+1}+Z\left(\tau-\sum_{k=j-P+1}^{j}T_{k}\right).
\end{align*}
Hence, for (\ref{eq:map-definition}) we obtain the form
\begin{equation}
T_{j+1}=1-Z\left(\tau-\sum_{k=j-P+1}^{j}T_{k}\right).\label{eq:T-OIOU}
\end{equation}

\subsection{Existence of RS}

In the case of RS the oscillator emits pulses periodically with $t_{j}=jT$
and $T_{j}=T$ for all $j$. This implies, that there is exactly one
spike within each interval $[t_{j},t_{j+1})$ and the ISI map necessarily
takes the form (\ref{eq:T-OIOU}).
\begin{figure*}
\begin{centering}
\includegraphics{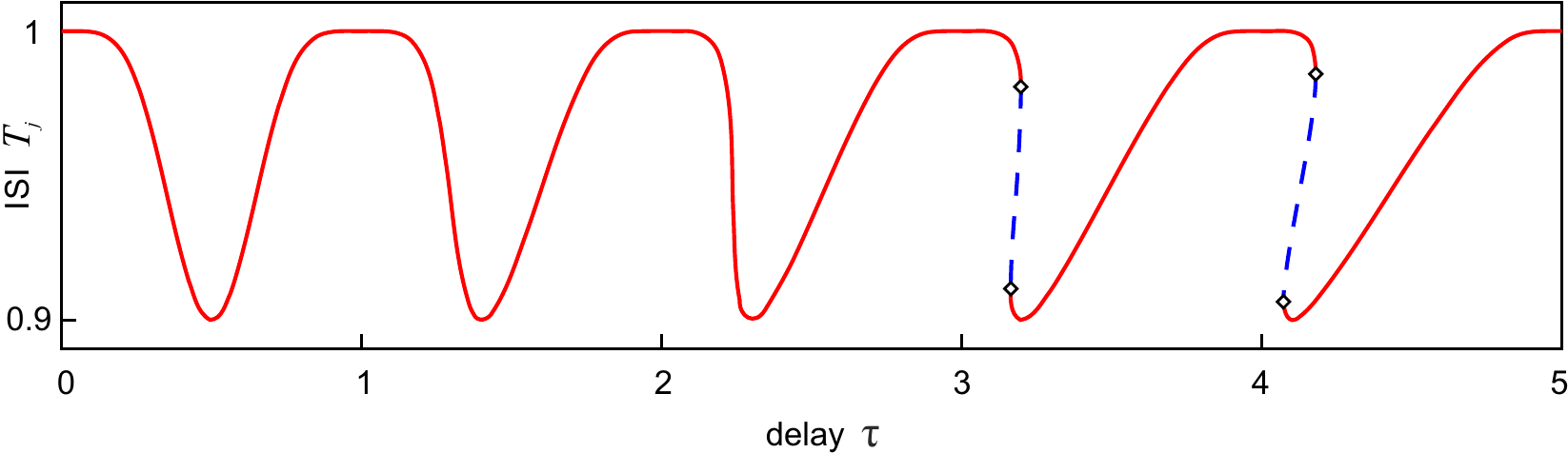}
\par\end{centering}

\protect\caption{\label{fig:b1d_q5} (Color online) One-dimensional bifurcation diagram
for the ISIs $T$ of RS solutions versus delay $\tau$ according to
(\ref{eq:T-par}) with $Z(\varphi)$ from Eq.~(\ref{eq:PRC}) and
$q=5$. Red solid lines correspond to stable RS, blue dashed lines
correspond to unstable RS {[}cf. Sec.~\ref{sec:Linear-stability}{]}.
The squares indicate fold bifurcations. }
\end{figure*}
 RS solutions correspond to fixed points of (\ref{eq:T-OIOU}) and
therefore all possible periods $T$ are given as solutions to
\begin{equation}
T=1-Z\left(\tau-PT\right)\label{eq:T-RS}
\end{equation}
where $P=\left[\tau/T\right]$ is the number of full periods within
one delay interval ($\left[\cdot\right]$ denotes the integer part).
Thus, $\tau=PT+\psi$ with $\psi:=\tau\mod T$, and we can write Eq.~(\ref{eq:T-RS})
in the parametric form: 
\begin{equation}
\begin{array}{l}
T=1-Z(\psi),\\
\tau=P\left(1-Z(\psi)\right)+\psi.
\end{array}\label{eq:T-par}
\end{equation}
The advantage of Eq.~(\ref{eq:T-par}) as compared to Eq.~(\ref{eq:T-RS})
is that it allows an explicit representation of all RS solutions (i.e.
their periods $T$) as a function of $\tau$. The curve $(\tau(\psi),T(\psi))$
obtained by substituting $P=0,1,2,...$ in (\ref{eq:T-par}) and varying
$\psi$ within the interval $[0,1]$ is shown in Fig.~\ref{fig:b1d_q5}.
An important consequence of this dependency is that for each value
of $\tau$ at least one value of $T$ exists. For small values of
$\tau$ the expression $T(\tau)$ is single-valued, but for larger
$\tau$ folding takes place leading to an emergence of several solution
branches. The fold points satisfy
\[
0=\frac{\partial\tau}{\partial T}=\frac{\partial\tau}{\partial\psi}\left(\frac{\partial T}{\partial\psi}\right)^{-1}=\frac{1-PZ^{\prime}\left(\psi\right)}{-Z^{\prime}\left(\psi\right)},
\]
that is,
\begin{eqnarray}
PZ^{\prime}(\psi) & = & 1.\label{eq_fold}
\end{eqnarray}
From Eq.~(\ref{eq_fold}) it is clear that folds do not occur for
$P<1/|\max_{0\le\varphi\le1}Z^{\prime}\left(\varphi\right)|$. For
larger $P$ intervals of $\tau$ appear for which several different
RS solutions with different periods coexist. As follows from a general
result derived in \citep{Yanchuk2009}, the number of coexisting RS
solutions grows linearly with the time-delay $\tau$.

\subsection{Linear stability\label{sec:Linear-stability}}

Let us analyze the linear stability of the RS solutions. For this
purpose we introduce small perturbations $\delta_{j}$ to the initial
conditions such that $T_{j}=T+\delta_{j}$, and study whether the
perturbations are damped or amplified with time. Since small perturbations
do not violate the property that only one spike occurs within each
interval $[t_{j},t_{j+1}]$, the map (\ref{eq:T-OIOU}) can be used
to study their evolution relevant to the stability of the corresponding
RS solution. We substitute $T_{j}=T+\delta_{j}$ into (\ref{eq:T-OIOU}),
linearize the obtained expression with respect to $\delta_{j}$, and
obtain the map
\begin{equation}
\delta_{j+1}=Z^{\prime}(\psi)\sum_{k=j-P+1}^{j}\delta_{k},\label{eq_seq_lin}
\end{equation}
where $\psi=\tau-PT$. The stability of the linear map (\ref{eq_seq_lin})
is described by its characteristic multipliers $\lambda\in\mathbb{C}$,
that is, by the roots of the characteristic equation
\begin{equation}
\chi_{P,\alpha}\left(\lambda\right)=\lambda^{P}-\alpha\sum_{k=0}^{P-1}\lambda^{k}=0\label{eq_char}
\end{equation}
where $\alpha:=Z^{\prime}(\psi)$. If Eq.~(\ref{eq_char}) has only
multipliers with $|\lambda|<1$ the corresponding RS regime is locally
stable. The following statements summarize properties of the multipliers
{[}see Appendix~\ref{sec:Appendix-spectrum} for details{]}:

(A) For $-1<\alpha<1/P,$ all multipliers have absolute values less
than one. As a result, the RS solution is asymptotically stable. 

(B) At $\alpha=1/P$ one critical multiplier crosses the unit circle
at $\lambda=1$. For $\alpha>1/P$ this multiplier remains unstable.

(C) At $\alpha=-1$ there are $P$ critical multipliers $\lambda_{k}=e^{i2\pi k/(P+1)},$
$k=1,\dots,P$, crossing $\left|\lambda\right|=1$ simultaneously.
For $\alpha<-1$, there are $P$ unstable multipliers with $|\lambda_{k}|>1$,
$k=1,\dots,P$. 

\begin{figure}
\begin{centering}
\includegraphics{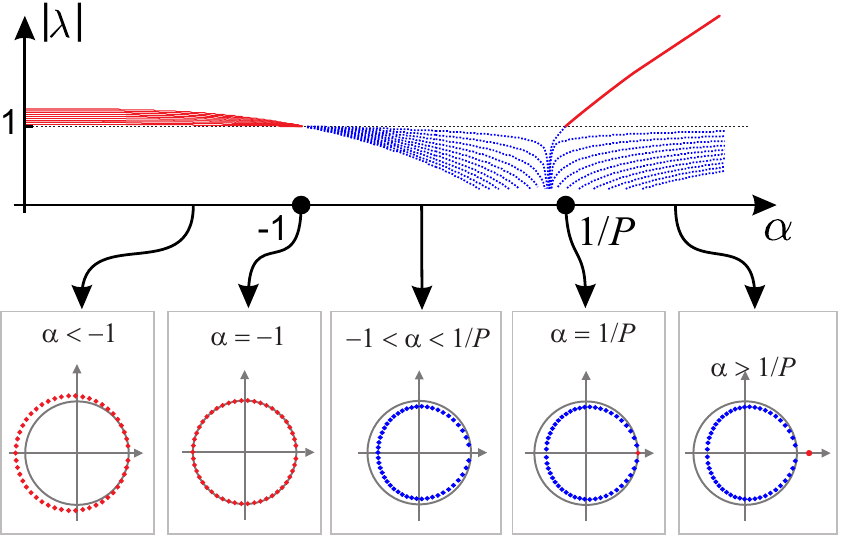} 
\par\end{centering}

\protect\caption{\label{fig:lambda}(Color online) Multipliers $\lambda$ of the RS
solution versus the parameter $\alpha$. Dotted blue lines depict
the stable part of the spectrum, solid red the unstable part. }
\end{figure}

Figure \ref{fig:lambda} illustrates the possible spectra for different
values of the parameter $\alpha.$ The stability region for regular
spiking is $-1<\alpha<1/P$. For $\alpha=1/P$ the stability is lost
through a saddle-node bifurcation corresponding to the fold described
by (\ref{eq_fold}).

The most remarkable destabilization scenario is related to the transition
at the parameter value $\alpha=-1$, where $P$ multipliers become
unstable simultaneously if $\alpha$ is decreased. At this point the
dimension of the unstable manifold increases abruptly from $0$ (stable
RS solution) to $P$, which can be arbitrary large depending on the
size of the delay $\tau$. Although this ``dimension explosion''
seems to be very degenerate, it occurs generically within our setup.
In the following we study this surprising bifurcation.

\section{Numerical study of jittering\label{sec:Numerical}}

In this section we show numerically, that a destabilization of RS
taking place at $\alpha=-1$ leads to the emergence of a variety of
``jittering'' spiking modes with non-identical ISIs $T_{j}$. In
order to achieve the bifurcation condition $\alpha=Z'(\psi)=-1$,
the steepness $s$ of the PRC should satisfy $s\ge1$. For the PRC
given by Eq.~(\ref{eq:PRC}), the steepness is given by (\ref{eq:steepness-formula}),
and the condition $s\ge1$ leads to $q\ge q^{\ast}$ with $q^{\ast}\approx e/(0.1\pi)^{2}\approx27.5$.

\begin{figure*}
\begin{centering}
\includegraphics{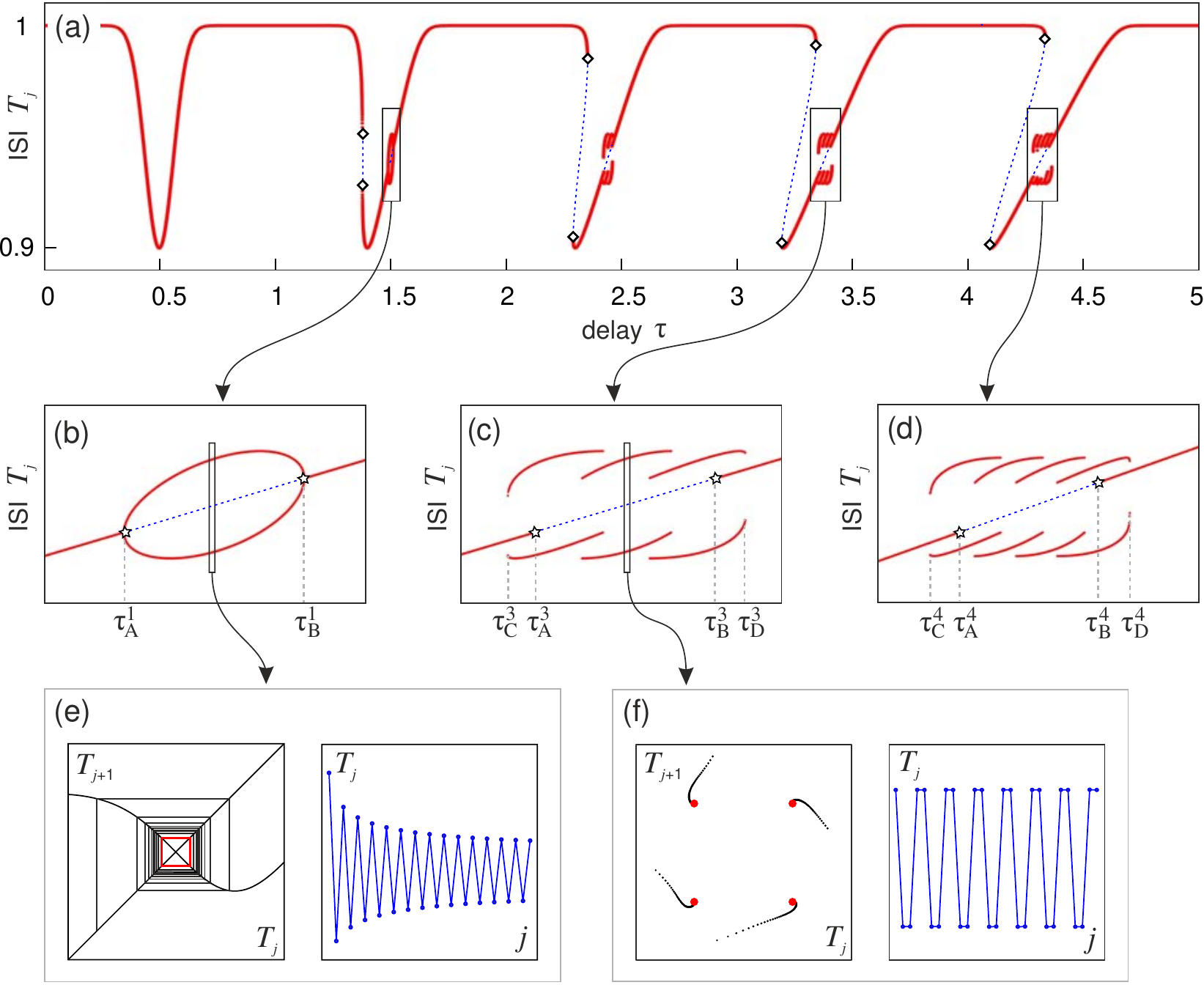}
\par\end{centering}

\protect\caption{\label{fig:b1d_q28} (Color online) (a) Numerical bifurcation diagram
for (\ref{eq:1}) with PRC (\ref{eq:PRC}) where $q=28$. The delay
$\tau$ is varied in the range $[0,5]$. Red dots correspond to ISIs
observed in direct simulations; the blue dashed lines corresponds
to unstable RS; squares indicate fold bifurcations. (b-d) are zooms
into regions of (a) where irregular spiking occurs; stars indicate
multi-jitter bifurcations. (e) Left panel: One-dimensional map (\ref{eq:T-Peq1})
describing the dynamics of ISIs for $P=1$ ($\tau=1.5$), together
with a cobweb diagram for a trajectory converging to the stable period-2
solution. Right panel: temporal dynamics, ISIs $T_{j}$ versus time.
(f) Bipartite period-4 solution $(\overline{T_{1},T_{1},T_{2},T_{2}})$
for $P=3$ ($\tau=3.38$). Left panel: a trajectory converging to
this solution in the $(T_{j+1},T_{j})$-plane; Right panel: temporal
dynamics. }
\end{figure*}

To explore the regimes of irregular spiking we first study the map
(\ref{eq:T-OIOU}) for the PRC (\ref{eq:PRC}) with $\kappa=0.1$.
A series of bifurcation diagrams is shown in Figs.~\ref{fig:b1d_q5},
\ref{fig:b1d_q28}, and, \ref{fig:b1d_q100}. They illustrate the
observed ISIs of the system for continuously varying delay $\tau$
and three different values, $q=5,$ $28$, and $100$, of the steepness
parameter. For each value of the delay $\tau$ we simulated the system
$20$ times starting from random initial conditions (initial ISIs
were drawn from a uniform distribution in $[0.9,1.0]$). For each
simulation all different values of ISIs $T_{j}$, which could be observed
after a transient, were saved. For each observed value $T_{j}$, a
red dot was placed at coordinate $(\tau,T_{j})$. Further, the parametric
representation (\ref{eq:T-par}) was utilized to draw a line corresponding
to RS solutions. It is solid red where the RS is stable and dashed
blue where it is unstable, as determined from the condition $\alpha\in(-1,1/P)$.

\subsection{Subcritical steepness}

Figure~\ref{fig:b1d_q5} shows the case $q=5<q^{\ast}$, where the
only possible bifurcations are folds of the RS. This means, for each
value of $\tau$, each single point $(\tau,T)$ on the bifurcation
diagram corresponds to one possible RS solution. Intervals of $\tau$
where two distinct values of $T$ appear correspond to different coexisting
RS solutions with distinct periods. The periods of the observed regular
spiking are determined analytically by (\ref{eq:T-par}). 

When the steepness of the PRC increases and $q$ exceeds the critical
value $q^{\ast}$, two points $\psi_{A,B}\in\left(0,1\right)$ appear
for which $Z^{\prime}(\psi_{A,B})=-1$. This means that for appropriate
values of the delay time $\tau$, such that $\psi=\tau\,\mathrm{mod}T$
equals either to $\psi_{A}$ or $\psi_{B}$, the stability of the
RS changes and $P$ multipliers cross the unit circle simultaneously,
where $P=[\tau/T]$. More precisely: for each $P\in\mathbb{N}$, there
exist two values of the delay {[}cf. (\ref{eq:T-par}){]},
\begin{equation}
\tau_{A,B}^{P}=P(1-Z(\psi_{A,B}))+\psi_{A,B},\label{eq:bif-points-dim-explosion}
\end{equation}
for which the dimension of the unstable manifold of the RS explodes
from 0 to $P$. 

\subsection{Moderate supercritical steepness\label{sub:Moderate-supercritical-steepness}}

Figure~\ref{fig:b1d_q28} illustrates the case close to criticality
with $q=28$, corresponding to $s\approx1.017$, and slightly larger
than the critical parameter value $q^{\ast}\approx27.5$. The newly
occurring bifurcations are located on the stable, upward part of the
RS branch and the panels \ref{fig:b1d_q28}(b-d) show enlarged neighborhoods
of the corresponding intervals $[\tau_{A}^{P},\tau_{B}^{P}]$ for
$P=1,3,4.$ 

As shown in Section~\ref{sec:Linear-stability}, $P$ multipliers
$\lambda_{k}=e^{i2\pi k/(P+1)}$ cross the unit circle simultaneously
at $\tau=\tau_{A,B}^{P}$. For the case $P=1$, this means only one
multiplier $\lambda=-1$ crosses the unit circle in the points $\tau=\tau_{A,B}^{1}$.
Note that in this case the map (\ref{eq:T-OIOU}) is one-dimensional
and has the form
\begin{equation}
T_{j+1}=1-Z\left(\tau-T_{j}\right).\label{eq:T-Peq1}
\end{equation}
In this case, the bifurcation is a supercritical period doubling giving
birth to a stable period-2 solution existing for $\tau$ in the interval
$[\tau_{A}^{1},\tau_{B}^{1}]$. For this solution the spiking regime
is ''jittering'': the ISIs $T_{j}$ are not equal anymore but they
form an alternating sequence with $T_{2j+1}=T_{1}$ and $T_{2j}=T_{2}$,
with $T_{1}\ne T_{2}$. The temporal dynamics of the ISIs for the
period-2 solution is illustrated in Fig.\ref{fig:b1d_q28}(e) together
with a corresponding cobweb-diagram for the one-dimensional map (\ref{eq:T-Peq1}).
In the bifurcation diagram Fig.~\ref{fig:b1d_q28}(a) a period-2
solution corresponds to a pair of points $(\tau,T_{1})$ and $(\tau,T_{2})$.

For $P\geq2$ bifurcations take place, where $P$ multipliers simultaneously
becoming unstable at $\tau=\tau_{A,B}^{P}$. The RS solution loses
its stability inside the interval $\tau\in\left[\tau_{A}^{P},\tau_{B}^{P}\right]$,
and various stable ``jittering'' regimes with non-equal ISIs appear.
In numerical studies we observe that the emerging solutions have period
$(P+1)$. However, a consistent property of these period-($P+1$)
solutions is that their ISIs consist of only two (or less often three)
different values of $T_{j}$. An example of such a period-4 solution
at $P=3$ is given in Fig. \ref{fig:b1d_q28}(f), where the corresponding
ISI sequence has the form $(\overline{T_{1},T_{1},T_{2},T_{2}}):=(\dots,T_{1},T_{1},T_{2},T_{2},\dots)$.
Here and in the following, the periodically repeating part of the
solution is denoted with an overline. As a result, such solutions
correspond to only two, and not $P+1$ points on the bifurcation diagram
in Fig.~\ref{fig:b1d_q28}(a). Since this type of solutions appear
to be prevalent in the considered system, we will introduce the term
\emph{bipartite} solutions to refer to them. Analogously, we use the
term \textit{tripartite} solutions for solutions exhibiting three
different ISIs.

For each $P\geq2$, a variety of different bi- or tripartite solutions
with period $P+1$ is observed inside the interval $\tau\in\left[\tau_{A}^{P};\tau_{B}^{P}\right]$.
Moreover, some of them exist in a wider parameter interval. We denote
this interval as $\left[\tau_{C}^{P};\tau_{D}^{P}\right]$. The stability
regions of different bipartite solutions alternate inside this interval,
so that different solutions are observed as the delay is changed.
For example, at $P=3$ the following period-4 bipartite solutions
are observed as $\tau$ changes from $\tau_{C}^{3}$ to $\tau_{D}^{3}$:\\
(i) $(\overline{T_{1},T_{1},T_{1},T_{2}})$,\\
(ii) $(\overline{T_{1},T_{2}})\equiv(\overline{T_{1},T_{2},T_{1},T_{2}})$,\\
(iii) $(\overline{T_{1},T_{1},T_{2},T_{2}})$,\\
(iv) $(\overline{T_{1},T_{2},T_{2},T_{2}})$,\\
where $T_{2}>T_{1}$. The stability regions of different solutions
overlap leading to multistability for the corresponding values of
the delay $\tau$. Tripartite solutions of the same period $P+1$
can be observed in a relatively narrow parameter interval at $P=2$
{[}cf. Sec.~\ref{sec:Solution_explosion}{]}.
\begin{figure*}
\centering{}\includegraphics{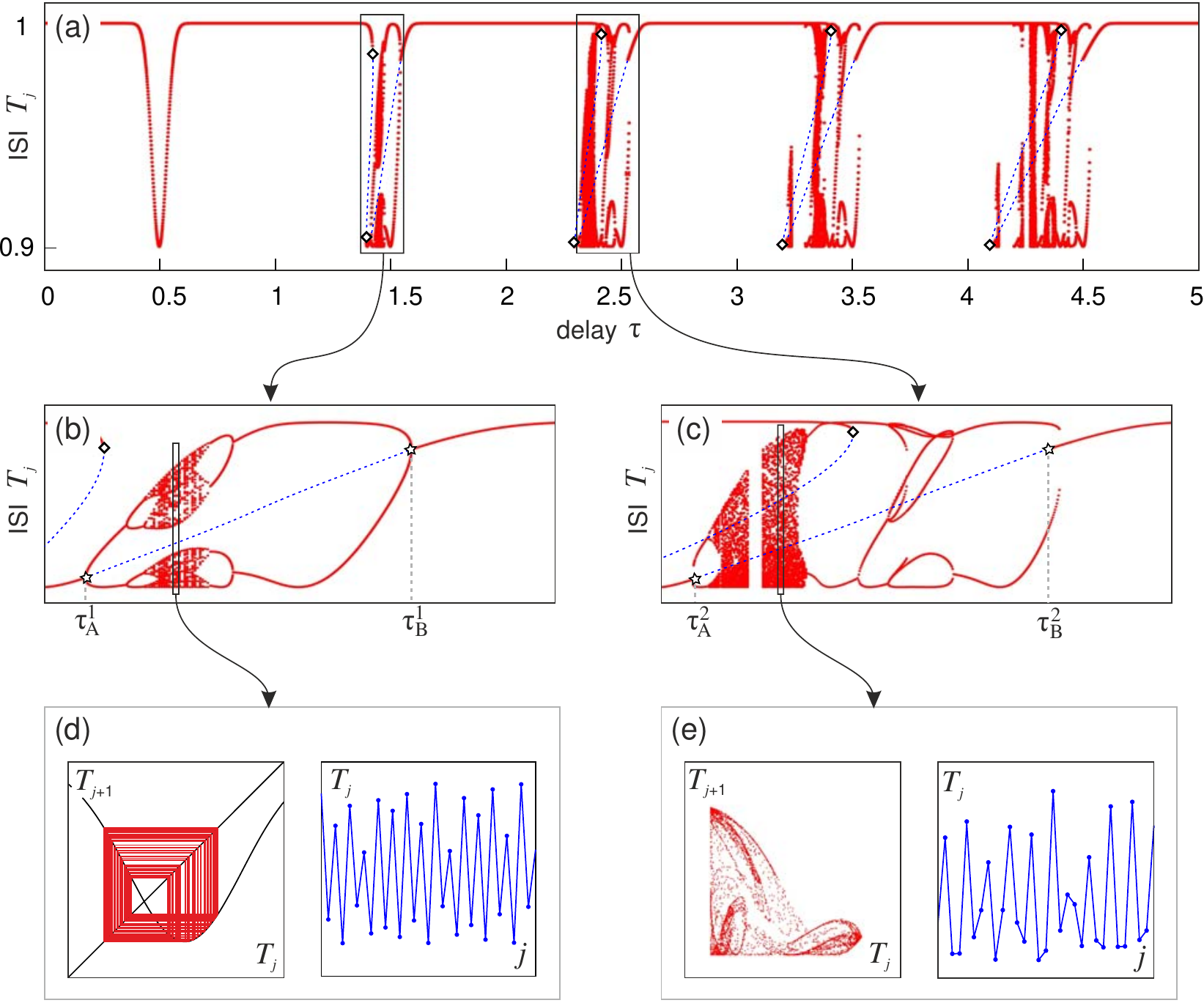}\protect\caption{\label{fig:b1d_q100}(Color online) (a) Numerical bifurcation diagram
for (\ref{eq:1}) with PRC (\ref{eq:PRC}) where $q=100$. The delay
$\tau$ is varied in the range $[0,5]$. Red dots correspond to ISIs
observed in direct simulations; the blue dashed lines correspond to
unstable RS; squares indicate fold bifurcations. (b) and (c) are zooms
of the indicated regions in (a); stars denote multi-jitter bifurcations.
(d) Chaotic solution for $P=1$ (right) and the corresponding 1-D
map (left) for $\tau=1.445$. (e) Chaotic solution for $\tau=2.38$
$(P=2)$ in the $(T_{j},T_{j+1})$-plane (left) and as a temporal
sequence (right). }
\end{figure*}

\subsection{Large supercritical steepness}

Figure~\ref{fig:b1d_q100} shows the case $q=100$, which is farther
beyond the criticality $q^{*}$ and corresponds to a steepness $s\approx1.9$.
As in the previous case, two points $\tau_{A,B}^{P}$ exist for each
$P=1,2,...$, where $\alpha=-1$, and bifurcations of the RS occur
at those points. The instability intervals $[\tau_{A}^{P},\tau_{B}^{P}]$
are larger than in Fig.~\ref{fig:b1d_q28} as well as the intervals
$\left[\tau_{C}^{P},\tau_{D}^{P}\right]$, where jittering regimes
appear. Besides the described bi- and tripartite solutions, more complex
dynamics is observed inside these intervals. For $P=1$ the scenario
of emergence of this dynamics is similar to that in the one-dimensional
logistic map. In this case a cascade of period doubling bifurcations
leads to the birth of a chaotic attractor {[}Fig.~\ref{fig:b1d_q100}(b){]},
which is illustrated in Fig.~\ref{fig:b1d_q100}(d). For $P\geq2$
the observed scenarios include period doubling cascades as well as
emergence and destruction of tori which both may lead to the birth
of chaotic attractors. An example of such an attractor for the case
$P=2$ is shown in Fig.~\ref{fig:b1d_q100}(e).
\begin{figure*}
\begin{centering}
\includegraphics{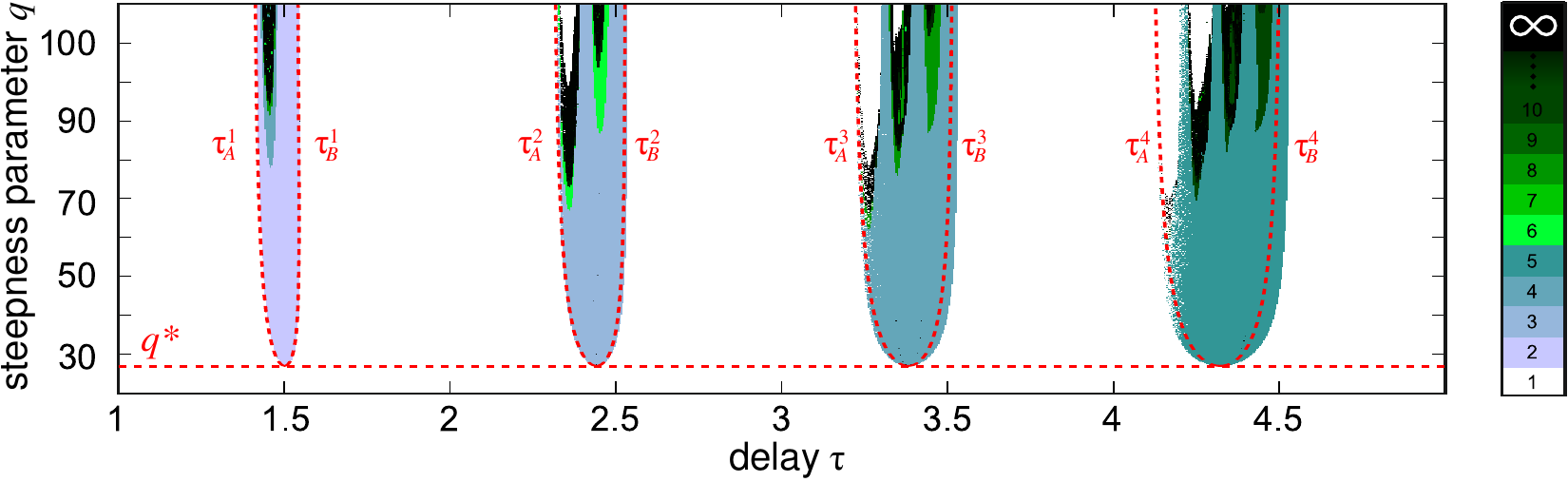}
\par\end{centering}

\protect\caption{\label{fig:bd2} (Color online) Numerically obtained two-dimensional
bifurcation diagram for system (\ref{eq:1}) with PRC (\ref{eq:PRC}).
Free parameters are the delay time $\tau$ and the steepness parameter
$q$. The maximal observed attractor period is coded in color. White
corresponds to period-1 (RS), shades of blue and green (gray) correspond
to finite numbers larger than one (bipartite and multipartite spiking),
black corresponds to quasiperiodic or chaotic spiking, or a period
larger than 100. Red dashed curves consist of multi-jitter bifurcations
points (corresponding to curves $\tau_{A,B}^{P}(q)$), and the horizontal
dashed red line indicates the critical steepness at $q=q^{*}$.}
\end{figure*}

A two-dimensional, numerical bifurcation diagram in the parameters
$\tau$ and $q$ is shown in Fig.~\ref{fig:bd2}. The diagram was
obtained by simulating (\ref{eq:1}) with $N=20$ different random
initial values (as for the one-dimensional diagrams in Figs.~\ref{fig:b1d_q28}
and \ref{fig:b1d_q100}) for each of $1000\times1000$ grid points
in the shown region $(\tau,q)\in[1,5]\times[20,120]$. For each point,
the period of the solution is computed after a transient for each
of the $20$ runs and the maximal observed period was stored to generate
the figure. If the period exceeded $100$, the observed regime was
considered aperiodic (black points).

The largest area (white) in Fig.~\ref{fig:bd2} corresponds to stable
RS solutions. For $q>q^{\ast}$, islands of irregularity appear, each
corresponding to one value of $P=1,2,3...$. The internal structure
of these islands is quite complicated. They consist of areas with
solutions of different periods, often connected via period doubling
bifurcations, as well as areas with quasiperiodic and chaotic solutions.
Close to the border of each island there are bipartite period-($P+1$)
solutions. Deeper in the interior, solutions of higher periods emerge,
as well as quasiperiodic and chaotic solutions. However, we also observe
windows of regularity inside the irregularity islands.

\section{The multi-jitter bifurcation\label{sec:Solution_explosion}}

As it was observed in Secs.~\ref{sec:Linear-stability} and \ref{sec:Numerical},
a RS solution is destabilized either in a saddle-node bifurcation
or in a peculiar bifurcation, where $P$ multipliers become unstable
simultaneously. In this section we show that a large number of ``jittering''
solutions emerges, i.e. solutions with different ISIs, in this ``multi-jitter''
bifurcation. To prove this we consider the equation
\begin{equation}
1-T=Z(T-\theta),\label{eq:theta}
\end{equation}
where $\theta>0$ is a constant and $T\in[\theta,\theta+1]$. As we
show below, solutions to this equation can be used as a basis for
the construction of periodic solutions of the map (\ref{eq:T-OIOU}). 

Let us first consider the simplest case, which assumes a single solution
$\tilde{T}$ of (\ref{eq:theta}). This implies immediately the existence
of RS with period $\tilde{T}$ for the delay values $\tau=\tilde{T}-\theta+P\tilde{T}$,
for all $P\in\mathbb{N}$ {[}cf. Eq.~(\ref{eq:T-RS}){]}. This observation
of multiple occurrence of the same solution at different values of
$\tau$ is well known as periodic solution reappearance in delay differential
equations \citep{Yanchuk2009}. 

Many more possibilities arise when for some value of $\theta$ Eq.~(\ref{eq:theta})
has two or more solutions $\tilde{T}_{k}.$ In this case one can construct
an arbitrary $(P+1)$-periodic sequence $(\overline{T_{1},T_{2},...,T_{P+1}})$
from these values, i.e. $T_{j}\in\left\{ \tilde{T}_{k}\right\} $,
and obtain a periodic solution of the map (\ref{eq:T-OIOU}) for
\begin{equation}
\tau=\left(\sum_{j=1}^{P+1}T_{j}\right)-\theta.\label{eq:delay-theta}
\end{equation}
This statement is readily confirmed by a direct check. Indeed, for
each item $T_{j+1}$ of the sequence $(\overline{T_{1},T_{2},...,T_{P+1}})$
we have
\[
\begin{array}{rl}
T_{j+1} & =T_{j-P}=1-Z(T_{j-P}-\theta)\\
 & =1-Z\left(\tau-\sum_{k=j-P+1}^{j}T_{j}\right),
\end{array}
\]
which coincides with (\ref{eq:T-OIOU}).

\begin{figure}
\begin{centering}
\includegraphics{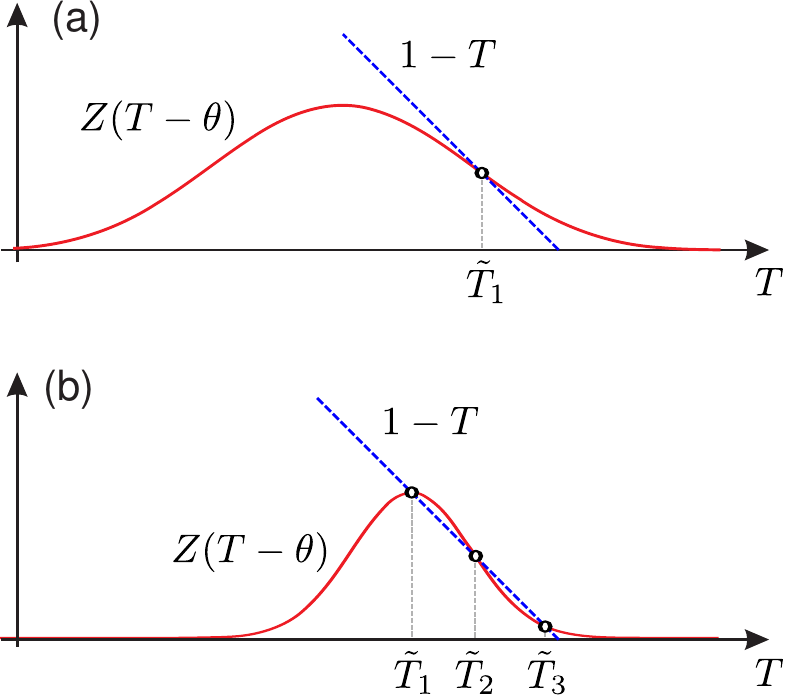}
\par\end{centering}

\protect\caption{\label{fig:theta}(Color online) Solutions of Eq.~(\ref{eq:theta})
for (a) $s<1$ and (b) $s>1$. The solid red line corresponds to the
right hand side, the dashed blue line to the left hand side of (\ref{eq:theta}).}
\end{figure}
Fig.~\ref{fig:theta} illustrates possible solutions to Eq.~(\ref{eq:theta})
for the PRC (\ref{eq:PRC}). If the slope of the right hand side,
that is of the PRC $Z$, is not less than $-1$ for all values of
$T$, only one intersection and only one solution $T_{1}$ of (\ref{eq:theta})
exists {[}Fig.~\ref{fig:theta}(a){]}. But if the PRC is steep enough,
namely $s>1$, three different solutions $T_{1}$, $T_{2}$ and $T_{3}$
of (\ref{eq:theta}) exist within a certain interval of $\theta$
(Fig.~\ref{fig:theta}(b)). In this case, the values of $\theta$
corresponding to the emergence of the new roots can be found from
the condition that the left hand and the right hand side of (\ref{eq:theta})
intersect tangentially in $\theta$. These values equal 
\[
\theta_{A,B}=1-\psi_{A,B}-Z(\psi_{A,B}),
\]
where $\psi_{A,B}$ are the points where the slope of the PRC equals
$Z^{\prime}\left(\psi_{A,B}\right)=-1$. Equation~(\ref{eq:theta})
has three different solutions inside the interval $\theta\in[\theta_{A},\theta_{B}]$
and only one solution outside of this interval. It can not have more
than three different solutions for the PRC given by (\ref{eq:PRC}). 

\begin{figure*}
\centering{}\includegraphics{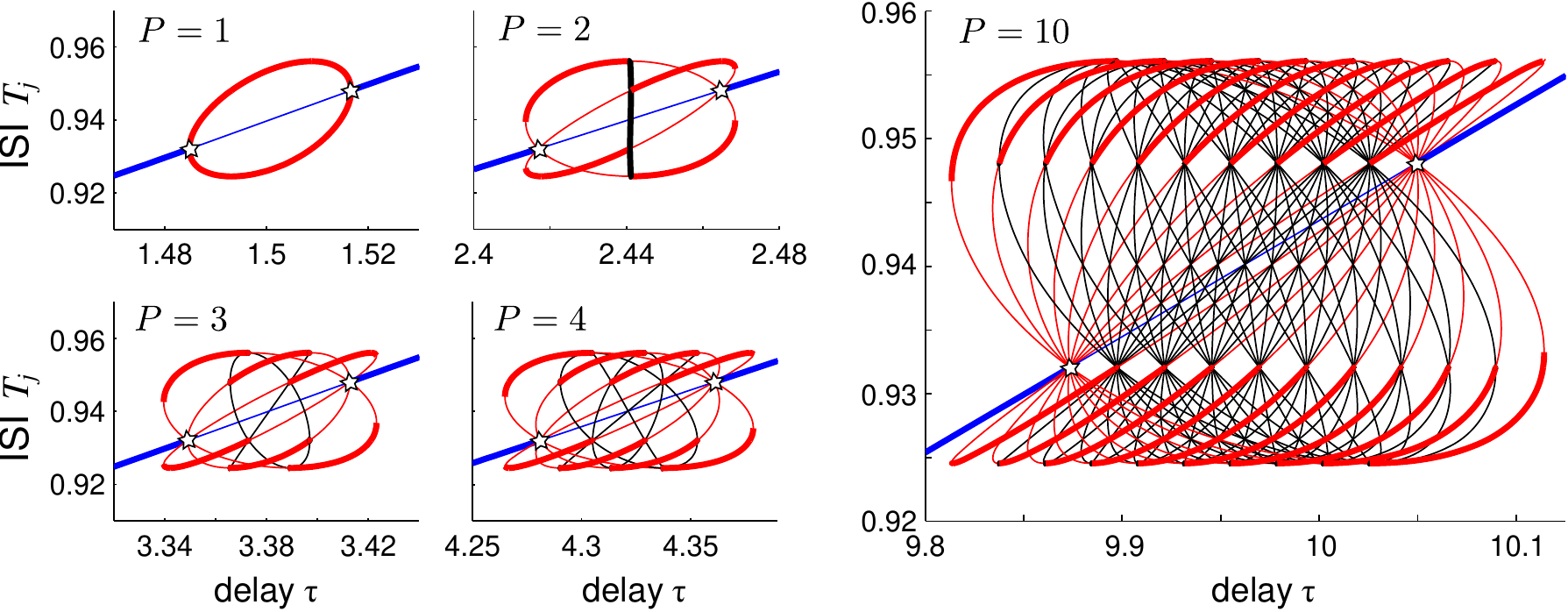}\protect\caption{\label{fig:elip}(Color online) (a)--(d) Branches of RS (blue or light
gray), bipartite (red or dark gray), and tripartite (black) solutions
in system (\ref{eq:1}) with PRC (\ref{eq:PRC}) and $q=28$, for
different values of $P$ as indicated in the plots. Stable parts of
the branches are shown by thick, and unstable by thin lines; stars
indicate multi-jitter bifurcations.}
\end{figure*}

This explains the emergence of bipartite and tripartite solutions
which were reported in Sec.~\ref{sec:Numerical}. They exist for
steepness $s>1$ when Eq.~(\ref{eq:theta}) can have more than one
solution $\tilde{T}_{k}$. To construct all solutions for a given
$P$ these solutions $\tilde{T}_{k}(\theta)$ are determined in dependence
of $\theta\in[\theta_{A},\theta_{B}]$ and then composed in all possible
periodic sequences of period $\le P+1$. A series of the bipartite
and tripartite solution branches obtained in this way is shown in
Fig.~\ref{fig:elip} for $q=28$ and $P=1,...,4$, and $P=10$. The
stability of these solutions was calculated as described in Appendix~\ref{sec:Stability-of-quasi-reappearant}. 

The obtained stable solutions coincide with the attractors from the
bifurcation diagrams in Fig.~\ref{fig:b1d_q28}(b)--(d) and complement
the diagrams by parts which are difficult to obtain by direct simulation
like unstable and tripartite solutions. Each observed bipartite solution
corresponds to a pair of points $(\tau,T_{1})$ and $(\tau,T_{2})$
on the diagram; each tripartite solution corresponds to three such
points. Note that solution branches which contain the same quantities
of each contained ISI coincide. For instance, in the case $P=3$ the
branches corresponding to bipartite solutions of the form $(\overline{T_{1},T_{2}})\equiv(\overline{T_{1},T_{2},T_{1},T_{2}})$
and $(\overline{T_{1},T_{1},T_{2},T_{2}})$ lie on top of each other
in Fig.~\ref{fig:elip}(c). This increasing number of overlapping
branches is the reason for the exponential growth of coexisting solutions
and not the number of visibly different branches which equals $P$.
Surprisingly, the stability along the overlapping branches seems to
coincide. We formulate this as a conjecture in Appendix~\ref{sec:Stability-of-quasi-reappearant}.
Notably, there exists only a very narrow interval of $\tau$ where
stable tripartite solutions exist. These are the solutions $(\overline{T_{1},T_{2},T_{3}})$
and $(\overline{T_{1},T_{3},T_{2}})$ for $P=2$. All other tripartite
solutions are unstable.

Let us now estimate the number of different bipartite solutions which
exist for a given $P$. Each of this solutions corresponds to a sequence
of two ISIs $T_{1}$ and $T_{2}$ which has the length $P+1$. For
each possibility to write $P+1=n_{1}+n_{2}$ with positive integers
$n_{1}$ and $n_{2}$, we obtain $\binom{P+1}{n_{1}}$ different sequences
consisting of $n_{1}$ entries equal to $T_{1}$ and $n_{2}$ entries
equal to $T_{2}$. Not all of these sequences correspond to different
periodic solutions of (\ref{eq:T-OIOU}), since some of them might
be transformable to others by a periodical shift. Both sequences correspond
to the same periodic solution if and only if this is possible. Therefore
we can estimate the total number of different $(P+1)$-periodic solutions
containing exactly $n_{1}$ ISIs $T_{1}$ in their sequential representation
to be equal to or larger than $\binom{P+1}{n_{1}}/(P+1)$. Here the
quotient $(P+1)$ disregards possible shift duplicates. Summing up
over $n_{1}=1,...,P$ gives the following estimate for the total number
of bipartite solutions existing for $P$:
\begin{equation}
\#\{\text{bipartite solutions for }P\}\geq\left(2^{P+1}-2\right)/(P+1).\label{eq:sol_number}
\end{equation}
Thus, the number of solutions emerging due to described mechanism
grows exponentially with $P$. Notice that these solutions exist for
different intervals of the delay $\tau$. However, all of them emerge
in the dimension explosion points $\tau_{A,B}^{P}$ and exist at least
in the interval $[\tau_{A}^{P},\tau_{B}^{P}]$. 

Indeed, consider a parametrized bipartite solution $(\overline{T_{1},...,T_{P+1}})$
with $T_{k}=T_{k}(\theta)$ such that, e.g. $T_{k}\left(\theta\right)\in\{\tilde{T}_{1}\left(\theta\right),\tilde{T}_{2}\left(\theta\right)\}$
{[}see Fig.~\ref{fig:theta}{]}. The value of $\theta$, say $\theta_{B},$
where both basic ISIs $\tilde{T}_{1}$ and $\tilde{T}_{2}$ converge
to the common value $\tilde{T}_{1,2}\to T=1-Z(\psi_{B})$ corresponds
to the point where the bipartite solution emerges from the branch
of RS solutions. At the same time the corresponding delay converges
to the value $\tau=P(1-Z(\psi_{B}))+\psi_{B}=\tau_{B}^{P}$. These
limits are exactly the values of the ISI and the delay in the points
of dimension explosion bifurcation of the RS as obtained from Eq.~(\ref{eq:bif-points-dim-explosion}). 

This finding is well recognizable by the diagrams in Fig.~\ref{fig:elip}
in which all branches of bipartite solutions start from the dimension
explosion points on the RS branch. The full branch of bipartite solutions
is obtained as the concatenation of the curves $\theta\mapsto(\tilde{T}_{2}(\theta),\tilde{T}_{3}(\theta))$,
$\theta\mapsto(\tilde{T}_{1}(\theta),\tilde{T}_{3}(\theta))$, and
$\theta\mapsto(\tilde{T}_{1}(\theta),\tilde{T}_{2}(\theta))$ with
$\theta$ running back and forth between $\theta_{A}$ and $\theta_{B}$.

Thus, we have shown that the ``dimension explosion'' of the unstable
manifold, which takes place at the multi-jitter bifurcation of the
RS regime is also a ``solution explosion''. In this bifurcation
numerous bipartite solutions branch off, and the number of the emergent
solution grows exponentially with the delay $\tau$. Since each of
these solutions corresponds to a jittering regime we adopt the term
``multi-jitter bifurcation''.

\section{Experimental observation of bipartite jittering solutions\label{sec:Experimental-observation-of}}

\begin{figure*}
\centering{}\includegraphics{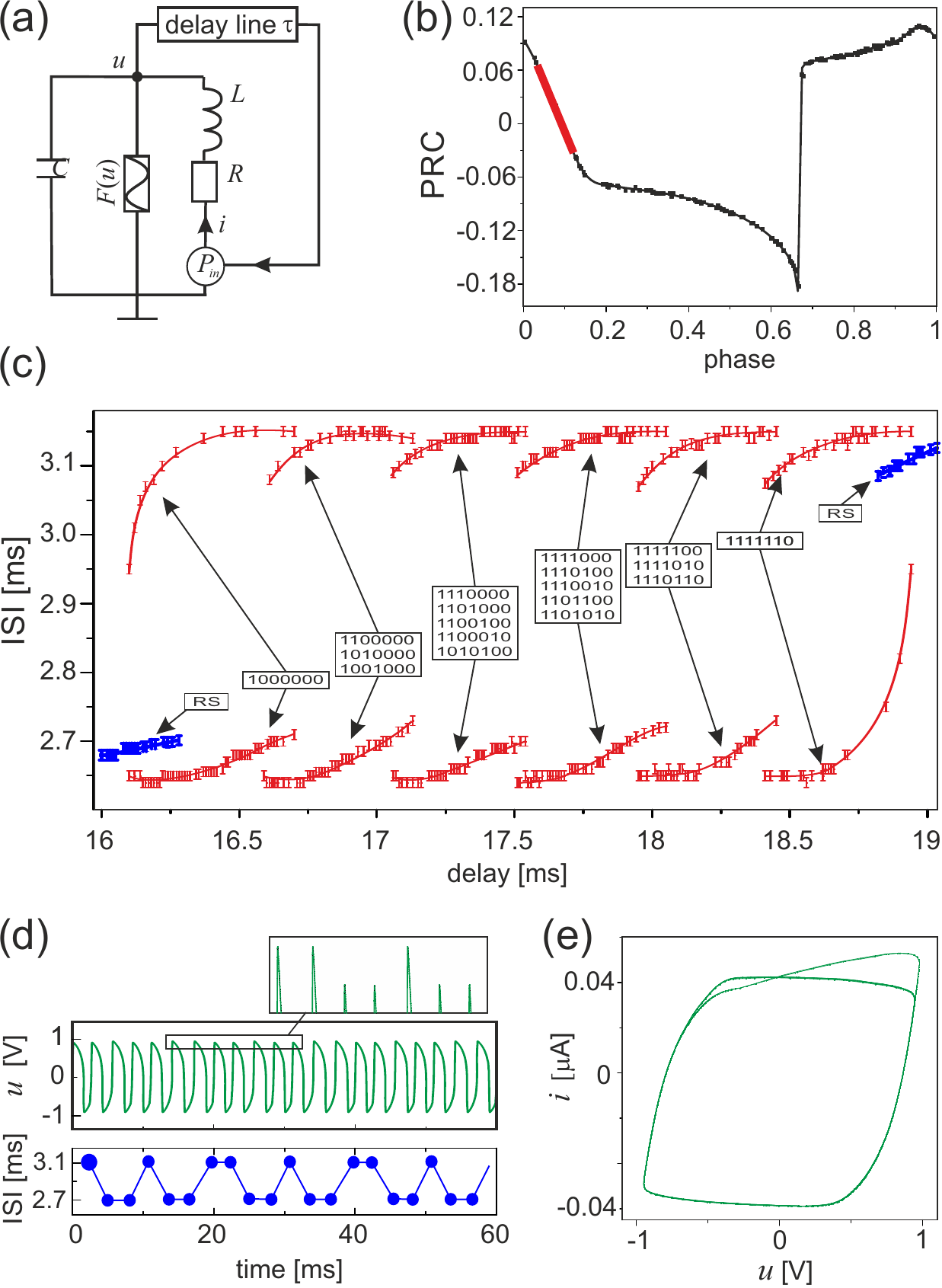}\protect\caption{\label{fig:exp}(Color online) Experimental study of jttering regimes
in an oscillatory electronic circuit. (a) Circuitry of the experimental
setup. (b) The measured PRC of the electronic oscillator. The thick
red line indicates the interval with slope < -1. (c) Bifurcation diagram
of the system for $P=6$: the observed ISIs are plotted against the
feedback delay time $\tau$. Blue (light gray) branches correspond
to regular spiking, red (dark gray) to jittering regimes. The binary
sequences in the inset boxes correspond to the observed jittering
solutions. (d) One of the observed jittering solutions: the dependence
of the output voltage $u$ and the ISI on the time. (e) The phase-plane
projection of the oscillator in the jittering regime.}
\end{figure*}

The multi-jitter bifurcation was discovered and studied in a reduced
phase model (\ref{eq:1}). An important question is whether this bifurcation
and the emerging multitude of jittering solutions is a peculiarity
of the phase models or it also occurs in realistic setups. 

In \citep{Klinshov2015a} we have provided some evidence that it also
occurs in a realistic neural model and in a physically implemented
electronic circuit. We have shown that the regular spiking destabilizes
in the points with $Z^{\prime}(\psi)=-1$ as predicted by our theory
and the jittering regimes emerge instead. However, it was not checked
whether all of the predicted jittering solutions can be realised.
According to (\ref{eq:sol_number}), their number should increase
exponentially with the delay.

We have experimentally studied an electronic FitzHugh-Nagumo oscillator
with a long feedback delay \citep{Shchapin2009,Klinshov2014}. The
circuitry of the electronic FitzHugh-Nagumo oscillator as used in
the experiment is depicted in Fig.~\ref{fig:exp}(a). Here, $R=1$k$\Omega$,
$C=5$nF, $L=9.4$H, $P_{in}$ is an input from the delay line, and
$F(u)=\alpha u(u-u_{0})(u+u_{0})$ is the current-voltage characteristic
of the nonlinear resistor with $\alpha=2.02\times10^{-4}\Omega^{-1}\mbox{V}^{-2}$
and $u_{0}=0.82$V. In absence of the delayed feedback, the device
exhibits autonomous oscillations with period $T\approx2.95$ms. The
delay line is realized on FPGA Xilinx Virtex-5 LX50 and it represents
a shift register consisting of 4000 elements with time of the shift
less than 7 microseconds. Thus, it provides a delay of a single pulse
with an accuracy better than 0.4\% of the autonomus oscillation period
T. The feedback is delivered as a pulse of amplitude $A=5$V and duration
$\theta=42\mu$s with a delay $\tau$ after each time the voltage
$u$ transverses the threshold $u_{th}=-0.7$V in positive direction.
For the given parameters, the phase resetting curve of the oscillator
has the shape depicted in Fig. \ref{fig:exp}(b) and exhibits an interval
with the slope less than $-1$ indicated by red color. 

During the experiment, we gradually selected different values of the
feedback delay time $\tau$, and recorded the distinct observed dynamical
regimes. The results are depicted in the experimental bifurcation
diagram in Fig.~\ref{fig:exp}(c). Here, for each delay $\tau$ the
observed ISIs are plotted analogously to the presentation in Figs.~\ref{fig:b1d_q5}--\ref{fig:b1d_q100}.
One can see that the regular spiking regime with a single value of
the ISI is observed for $\tau<\tau_{A}\approx16.3$ms and for $\tau>\tau_{B}\approx18.8$ms.
Inside the interval $[\tau_{A};\tau_{B}]$, the regular spiking regime
destabilizes and bipartite jittering solutions are observed. Each
bipartite solution corresponds to a pair of points on the bifurcation
diagram.

Recall that our theory predicts the coexistence of different bipartite
regimes with the same number of short and long ISIs differing only
in their order. In the experiment, we have considered values of the
delay within the range $\tau\in[16{\rm ms},19{\rm ms}]$ which corresponds
to $P=6$. According to the estimate (\ref{eq:sol_number}) at least
18 different jittering regimes should exist for the considered values
of $\tau$. Each of this regimes can be encoded by a binary sequence
of length $P+1=7$ representing a short ISI by zero and a long by
one. Not all of these regimes are distinguishable in the bifurcation
diagram since some of them are constituted of the same ISIs and thus
correspond to the same points. To detect experimentally the coexistence
of such solutions we temporarily applied an external noisy signal
to induce switches between different attractors. An example of a jittering
regime observed is shown in Fig.~\ref{fig:exp}(d). The solution
consists of a periodic repetition of four short and three long ISIs.
In the phase-plane projection of the system two loops are present
corresponding to two ISIs (Fig.~\ref{fig:exp}(e)).

In Fig.~\ref{fig:exp}(c), the observed bipartite regimes along each
depicted branch are indicated by the corresponding binary sequences.
As predicted by the theory, the branches with the same number of short
and long ISIs coincide. Moreover, the delay intervals in which such
branches are observed coincide as well. This suggests that, also for
the experimental system, the stability of a bipartite regime does
only depend on the number of short and long ISIs, not on their order.

\section{Discussion\label{sec:Discussion}}

In a phase oscillator with pulsatile feedback (\ref{eq:1}) the destabilization
of the regular spiking mode can occur through a peculiar bifurcation
in which the dimension of the unstable manifold ``explodes'', i.e.
changes abruptly from zero to an arbitrary large value. To our opinion,
it is remarkable and surprising that the single (and robust) condition
$Z^{\prime}\left(\varphi\right)<-1$ induces a destabilization of
the regular spiking regime along many directions simultaneously. Normally,
bifurcations in which many multipliers become critical simultaneously
have large codimension. This means that they take place in low-dimensional
subsets of a high-dimensional parameter space. In this sense, the
multi-jitter bifurcation in system (\ref{eq:1}) has codimension one,
since it occurs when a scalar equation of the system parameters is
fulfilled, namely $Z^{\prime}\left(\varphi\right)=-1$. Varying a
single system parameter, e.g. the delay, one can trigger this bifurcation.
Thus, it is a degenerate bifurcation which emerges generically. This
paradox originates in the structure of the map (\ref{eq:T-OIOU})
considered as a $P-$dimensional map
\begin{align}
\boldsymbol{T}=\left(T_{1},...,T_{P}\right)\nonumber \\
\mapsto\boldsymbol{T}^{{\rm new}} & =\left(T_{1}^{{\rm new}},...,T_{P}^{{\rm new}}\right)\label{eq:p-dim-map}\\
 & =\left(T_{2},...,T_{P},1-Z(\psi)\right)\nonumber 
\end{align}
where $\psi=\tau-\sum_{k=1}^{P}T_{k}$. Besides the calculation of
the new ISI $T_{P}^{{\rm new}}=1-Z(\psi)$ the map only shifts the
past ISIs by one component, which corresponds to the introduction
of a comoving timeframe. This results in a specific structure of the
map Jacobian
\begin{equation}
D(\alpha)=\left[\begin{matrix}0 & 1 & 0 & 0\\
\vdots & \ddots & \ddots & 0\\
0 & \cdots & 0 & 1\\
\alpha & \cdots & \cdots & \alpha
\end{matrix}\right],\label{eq:Jacobian}
\end{equation}
where $\alpha=Z^{\prime}\left(\psi\right)$. The matrix $D(\alpha)$
is the companion matrix of the characteristic equation (\ref{eq_char}).
Although one may think of arbitrary perturbations of the map (\ref{eq:p-dim-map})
only few of them are physically meaningful. For example, if we introduce
an additional parameter $\varepsilon$ to alter the first equation
of the map such that 
\[
T_{1}^{{\rm new}}=T_{2}+\epsilon g(\boldsymbol{T}),
\]
with some function $g:\mathbb{R}^{P}\to\mathbb{R}$, this would not
correspond to a reasonable perturbation because intervals of time
do not become longer or shorter while the origin of time is shifted
by a comoving timeframe. Similar reasoning reveals that the structure
of the whole map should be preserved. It seems that physically admissible
perturbations can only affect the delay $\tau$ or the PRC shape and
all these perturbations exclusively affect the coefficient $\alpha$
in the Jacobian (\ref{eq:Jacobian}) while its structure preserves.
Thus, the characteristic equation remains of the form (\ref{eq_char})
which implies a multi-jitter bifurcation for $\alpha=-1$, where many
multipliers become unstable at once.

Besides the discovery of the dimension explosion phenomenon, which
is surprising \emph{per se}, we proved that each such bifurcation
is accompanied by a potentially huge number of simultaneously emerging
jittering solutions. This observation lead us to adopt the name ``multi-jitter''
bifurcation. More precisely, the number of the emergent coexisting
solutions is exponential in the length of the delay $\tau$. This
phenomenon is akin to the reappearance of periodic solutions, a well-known
property of delay differential equations \citep{Yanchuk2009}. Consider
a $T$-periodic solution $x\left(t\right)=x\left(t+T\right)$ of the
general equation 
\[
\frac{dx}{dt}\left(t\right)=f\left(x\left(t\right),x\left(t-\tau\right)\right).
\]
Then it is clear from the periodicity of the solution $x\left(t\right)$
that it also solves the equation 
\[
\frac{dx}{dt}\left(t\right)=f\left(x\left(t\right),x\left(t-\tau-PT\right)\right)
\]
for any $P\in\mathbb{N}$. It means, that the same periodic solution
reappears at the delay times $\tau_{P}=\tau+PT$. Note that the stability
of the reappearing solution can be different for different delay times
(see more details in \citep{Sieber2013}). The number of coexisting
reappearant solutions increases linearly with the delay time, which
can be accessed intuitively from the observation that the reappearing
branch of solutions is stretched proportionally to the reappearance
index $P$ and the change in period along the branch {[}cf. the branch
of RS solutions in Fig.~\ref{fig:b1d_q5}{]}.

In contrast to this type of reappearance, the reappearant solutions
in system (\ref{eq:1}) may not be identical to the original solution,
but the possibility of ISI reordering allows for the emergence of
a multitude of new solutions. Thus, we observe reappearance not of
whole solutions, but of individual ISIs, which leads us to propose
to call the phenomenon ``quasi-reappearance''. For instance, the
existence of a period-2 solution with the ISIs $T_{1}$ and $T_{2}$
implies the existence of infinitely many bipartite solutions which
exhibit exactly these ISIs for larger delays. In fact, \emph{every}
periodical sequence of these two ISIs will correspond to a solution
of (\ref{eq:1}) for an appropriate value of the delay. Indeed, if
the sequence $(\overline{T_{1},T_{2}})$ solves (\ref{eq:T-OIOU})
for some $\tau=\tau_{0}$, and arbitrary periodical sequence of the
two ISIs containing $n_{1}$ instances of $T_{1}$ and $n_{2}$ instances
of $T_{2}$ per period solves (\ref{eq:T-OIOU}) for 
\begin{equation}
\tau=\tau_{0}+\left(n_{1}-1\right)T_{1}+\left(n_{2}-1\right)T_{2}.\label{eq:quasi-reapp-delay}
\end{equation}
The combinatorial variety of the quasi-reappearant solutions causes
an exponential increase of their number with the delay. Therefore
the multistability of the system increases potentially much faster
than due to ordinary reappearance.

An interesting property following from this mechanism of quasi-reappearance
is that the $\tau$-intervals for the existence of the jittering regimes
at a certain $P$ are ordered according to the corresponding number
of short/long ISIs {[}see, e.g., Fig.~\ref{fig:exp}(c){]}. The regime
with just \textit{one} long ISI is stable for the smallest values
of $\tau$ in the region corresponding to $P$. Subsequently, the
three regimes with \textit{two} long ISIs stabilize for larger $\tau$,
then the regimes with \textit{three} long ISIs stabilize and so on.
It is also noteworthy that the intervals of stability of the regimes
with $n_{1}$ and $n_{1}+1$ long ISIs may overlap giving rise to
even greater multistability. This feature is also confirmed by the
experiment: for example, for $\tau=17.52$ms, ten different jittering
regimes are observed {[}cf. Fig.~\ref{fig:exp}(c){]}.

We would also like to comment on the relevance of the multi-jittering
phenomenon to more realistic setups, where the oscillator's phase
space has a higher dimension than one and the pulse is smooth and
of finite duration. We have demonstrated that the multi-jitter instability
can appear in real electronic circuit as well as in a simulated Hodgkin-Huxley
neuron model \citep{Klinshov2015a} and it might be one of the mechanisms
behind the appearance of irregular spiking in neuronal models with
delayed feedback \citep{Ma2007} and timing jitter in semiconductor
laser systems with delayed feedback \citep{Otto2012} reported by
other authors. 

Our theory predicts and our numerical simulations confirm that the
essential property for the occurrence of a dimension explosion bifurcation
is the steepness of the PRC corresponding to the pulsatile feedback
action. The presence of regions where the slope of the PRC fulfills
$Z^{\prime}\left(\varphi\right)<-1$ requires some specific organization
of the oscillator's phase space. Consider the phase transition map 

\begin{equation}
f(\varphi)=\varphi+Z(\varphi),\label{eq:phase-transition-map-1-1-1}
\end{equation}
which describes change of the oscillator's phase under the action
of the feedback pulse. Notice that $Z^{\prime}\left(\varphi\right)<-1$
for some $\varphi$ if and only if the phase transition map is non-monotonous.
This means that the input can inverse the phase order of some points
from the limit cycle, such that $f(\varphi_{2})<f(\varphi_{1})$ for
some $\varphi_{2}>\varphi_{1}$.

\begin{figure}
\centering{}\includegraphics{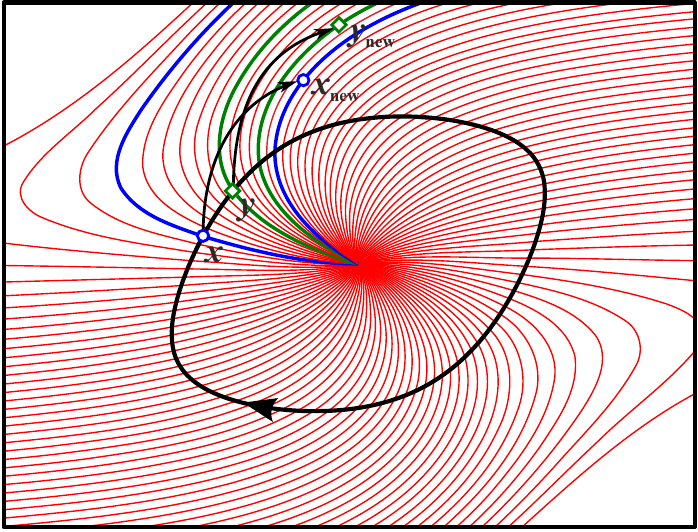}\protect\caption{\label{fig:isochrons-1}(Color online) The isochrons (thin red lines)
of the limit cycle (thick black) of a FitzHugh-Nagumo model show the
structure necessary for $Z^{\prime}\left(\varphi\right)<-1$. The
phase order of two different points on the limit cycle, $x$ and $y$,
is reversed by the corresponding pulse. This means they are carried
to new points $x_{{\rm new}}$ and $y_{{\rm new}}$ which lie on isochrons
with a reversed phase order.}
\end{figure}

This feature can be interpreted geometrically as a property of the
isochrons in the oscillator's phase space. The isochrons are sets
consisting of points that have the same phase, i.e., their distance
vanishes for $t\to\infty$ as they are attracted to the limit cycle
\cite{Guckenheimer1975,Winfree2001,isochrons1,isochrons2}. The new
phase $\varphi_{{\rm new}}=f(\varphi)$ induced by a pulse, which
starts when the oscillator is located at phase $\varphi$ on the limit
cycle, is determined by the isochron to which the point with the phase
$\varphi$ is conveyed by the pulse. The fact that the input inverses
the phase order of some neighboring points from the limit cycle implies
that the point of smaller phase transverses more isochrons than the
other. If the direction of the pulse does not change, which typically
leads to a straight displacement of the oscillator's state \cite{Kazantsev04,Klinshov08},
the isochrons should have a specific form of nested u-shaped curves.
This is illustrated in Fig.~\ref{fig:isochrons-1} for the FitzHugh-Nagumo
system.

The above reasoning shows that the multi-jitter bifurcation is possible
only for oscillators with specific structure of the phase space, namely
the limit cycle isochrons. However, there is a very easy criterion
reflecting this structure, which is the presence of regions with slope
< -1 in the  PRC.  PRCs possessing this property have been experimentally
measured here for the electronic oscillator and by other authors for
chemical oscillators \citep{Proskurkin2015,proskurkin2015new} and
cardiac cells \citep{Jalife1976,jalife1979phase,Guevara1981}. As
for neurons, PRCs with steep falling parts have been observed \citep{tsubo2007layer},
but there is no data whether their slope can be less than -1 or not.
Interestingly, it turns out that PRCs of both type I and type II with
slope $<-1$ can be easily obtained for many neural models \cite{Achuthan2009,Klinshov2015a,Schultheiss2010,krogh2012phase}.
This fact draws attention to the importance of assessing the slopes
of the PRCs measured for living neurons.

Thus, the discovered dimension explosion and multi-jittering phenomena
are relevant not only for a simple phase-reduced model, but for realistic
oscillatory systems as well. A promising application for systems exhibiting
high-dimensional complex dynamics is their utilization as liquid state
machines. For lasers with long delayed feedback this has been done
successfully \citep{Appeltant2011}. In view of the arbitrarily high
dimensional critical manifolds of the RS solution, which were observed
in this paper, it seems to be worth studying such computational abilities
for an oscillator with delayed pulsatile feedback. An electronic implementation
of such systems is relatively uncomplicated \citep{Shchapin2009,Klinshov2014,Klinshov2015a}.

\section{Conclusions\label{sec:Conclusion}}

In this paper we have studied the dynamics of oscillators with pulsatile
delayed self-feedback where we directed our attention to the influence
and interplay of the feedback delay and the shape of the oscillator's
PRC. As a major result we have proven that, if the oscillator's PRC
possesses sufficiently steep parts, the system undergoes a sequence
of degenerate bifurcations as the feedback delay is increased. These
multi-jitter bifurcations lead to a destabilization of the regular
spiking regime through the emergence of an unstable manifold of arbitrarily
large dimension, which is proportional to the delay size. 

Moreover, we proved that each multi-jitter bifurcation is accompanied
by emergence of numerous ``jittering'' periodic solutions with distinct
ISIs. We showed that the mechanism of emergence of such solutions
generalizes what is known as reappearance of periodic solutions. In
this \textquotedblleft quasi-reappearance\textquotedblright , individual
ISIs of solutions existing for smaller values of the delay reappear.
Thus, when the delay is increased the number of different solutions
grows combinatorially causing high multistability and extremely complex
structure of the system's phase space.
\begin{acknowledgments}
The theoretical study was supported by the Russian Foundation for
Basic Research (Grants No. 14-02-00042 and No. 14-02-31873), the German
Research Foundation (DFG) in the framework of the Collaborative Research
Center SFB 910, Erasmus Mundus Aurora consortium (grant number AURORA2012B1246),
and the European Research Council (ERC-2010-AdG 267802, Analysis of
Multiscale Systems Driven by Functionals). The experimental study
was carried out with the financial support of the Russian Science
Foundation (Project No. 14-12-01358).
\end{acknowledgments}

\appendix

\section{Steepness of (\ref{eq:PRC})\label{sec:Appendix-Steepness}}

Here we estimate the steepness {[}cf.~(\ref{eq:steepness-defenition}){]}
of the function $Z(\varphi)=\kappa\left(\sin\left(\pi\varphi\right)\right)^{q}$
for $\kappa>0$ and $q\gg1$. The slope equals
\[
Z'(\varphi)=\kappa\pi q\left(\sin\left(\pi\varphi\right)\right)^{q-1}\cos\left(\pi\varphi\right).
\]
It is extremal for $\varphi^{\ast}\in(0,1)$ with
\begin{align*}
 & Z''(\varphi^{\ast})\\
= & \kappa\pi^{2}q(q-1)\left(\sin\left(\pi\varphi^{\ast}\right)\right)^{q-2}\left(\cos\left(\pi\varphi^{\ast}\right)\right)^{2}\\
 & -\kappa\pi^{2}q\left(\sin\left(\pi\varphi^{\ast}\right)\right)^{q}\\
= & \kappa\pi^{2}q\left(\sin\left(\pi\varphi^{\ast}\right)\right)^{q-2}\left(q\left(\cos\left(\pi\varphi^{\ast}\right)\right)^{2}-1\right)=0
\end{align*}
In this point, we have
\begin{align*}
\cos\left(\pi\varphi^{\ast}\right) & =\pm\sqrt{\frac{1}{q}},\text{ and }\sin\left(\pi\varphi^{\ast}\right)=\pm\sqrt{1-\frac{1}{q}}\approx\pm(1-\frac{1}{2q}).
\end{align*}
Hence, the steepness can be calculated as
\begin{align*}
s & =|Z'(\varphi^{\ast})|=|\kappa\pi q\left(\sin\left(\pi\varphi^{\ast}\right)\right)^{q-1}\cos\left(\pi\varphi^{\ast}\right)|\\
 & \approx\kappa\pi q\left(1-\frac{1}{2q}\right)^{q-1}\frac{1}{\sqrt{q}}\approx\kappa\pi\sqrt{\frac{q}{e}},
\end{align*}
where $e$ is Euler's number.

\section{RS Spectrum\label{sec:Appendix-spectrum}}

Here we study the roots of the characteristic equation (\ref{eq_char}).
For convenience, we restate the properties of the spectrum asserted
in Section~\ref{sec:Linear-stability}:

(A) For $-1<\alpha<1/P,$ all multipliers $\lambda_{k}$, $k=1,...,P$,
have absolute value less than one.

(B) At $\alpha=1/P$ a critical multiplier crosses the unit circle
at $\lambda=1$. For $\alpha>1/P$ this multiplier remains unstable.

(C) At $\alpha=-1$ there are $P$ critical multipliers $\lambda_{k}=e^{i2\pi k/(P+1)},$
$k=1,\dots,P$, crossing $\left|\lambda\right|=1$ simultaneously.
For $\alpha<-1$, there are $P$ unstable multipliers with $|\lambda_{k}|>1$,
$k=1,\dots,P$. 

To prove these claims, we multiply the characteristic polynomial $\chi_{P,\alpha}\left(\lambda\right)$
by $(\lambda-1)$ and study the extended characteristic equation
\begin{align}
\tilde{\chi}_{P,\alpha}\left(\lambda\right) & =\left(\lambda-1\right)\chi_{P,\alpha}\left(\lambda\right)\nonumber \\
 & =\lambda^{P+1}-(1+\alpha)\lambda^{P}+\alpha=0.\label{eq_char_1}
\end{align}
The set $\tilde{\Lambda}$ of roots of $\tilde{\chi}_{P,\alpha}\left(\lambda\right)$
contains all roots $\Lambda=\{\lambda_{1},...,\lambda_{P}\}$ of $\chi_{P,\alpha}\left(\lambda\right)$
and the root $\lambda_{P+1}=1$, i.e., $\Lambda=\tilde{\Lambda}\backslash\{1\}$
for $\alpha\ne1/P$. In the following, we study the critical roots
of $\tilde{\chi}_{P,\alpha}\left(\lambda\right)$ , i.e. solutions
of (\ref{eq_char_1}) with $\left|\lambda\right|=1$. Substituting
$\lambda=e^{i\varphi}$ into (\ref{eq_char_1}) we obtain
\begin{equation}
e^{i\left(P+1\right)\varphi}+\alpha=(1+\alpha)e^{iP\varphi}.\label{eq:char-eq-for-critical-mult}
\end{equation}
Taking the absolute value on both sides of (\ref{eq:char-eq-for-critical-mult})
yields
\[
\left|e^{i\left(P+1\right)\varphi}+\alpha\right|=\left|1+\alpha\right|,
\]
which, for $\alpha\ne0$, implies $e^{i(P+1)\varphi}=1$. This means
$\varphi=2\pi k/(P+1)$ for some $k\in\mathbb{Z}$. Substituting this
into (\ref{eq:char-eq-for-critical-mult}) gives
\[
1+\alpha=(1+\alpha)e^{i2\pi kP/(P+1)}.
\]
For $\alpha\ne-1$, this requires $k\in(P+1)\mathbb{Z}$. Thus, for
$\alpha\notin\{-1,0\}$, $\lambda=1$ is the only solution of (\ref{eq_char_1})
with $\left|\lambda\right|=1$. It corresponds to a critical multiplier
of (\ref{eq_char}) only for $\alpha=1/P$, where it is a double root
of (\ref{eq_char_1}). Indeed, substituting $\lambda=1$ into (\ref{eq_char})
one obtains $1-P\alpha=0$. For $\alpha=-1$, (\ref{eq_char_1}) reduces
to
\[
\lambda^{P+1}=1.
\]
Hence, $P$ critical multipliers $\lambda_{k}=e^{i2\pi k/(P+1)},$
$k=1,...,P$, appear simultaneously at $\alpha=-1$. For $\alpha=0$,
(\ref{eq_char}) reduces to $\lambda^{P}=0,$ which obviously exhibits
no critical multipliers.

Finally let us remark that, with respect to $\alpha$, all critical
roots $\lambda\left(\alpha\right)$ of $\chi_{P,\alpha}$ traverse
the unit cycle at criticality $\left|\lambda\left(\alpha\right)\right|=1$
and for $\alpha=0$ all multipliers vanish identically: $\lambda_{1}\left(0\right)=...=\lambda_{P}\left(0\right)=0$.
This completes the prove of (A)--(C).

For $P\gg1$ a deeper study of the multipliers of (\ref{eq_char})
is possible. One may follow the concept developed in \citep{Heiligenthal2011}
and determine the so-called ``strong'' and ``weak'' multipliers.
Strong multipliers $\lambda_{s}$ are characterized by the fact that
$\lim\limits _{P\to\infty}\lambda_{s}(P)=\lambda_{s}$. The only strong
multiplier of (\ref{eq_char}) equals $\lambda_{s}\approx1+\alpha$
for large $P$, which means that the strong spectrum is unstable for
$\alpha>0$ and $\alpha<-2$. Weak multipliers $\lambda_{w}$ are
characterized by the following asymptotic behavior for large $P$:
$\lambda_{w}(P)\approx e^{\mu/P+i\omega}$ . The weak multipliers
of (\ref{eq_char}) are given by the relationship
\begin{equation}
\mu=-\frac{1}{2}\ln\left(1+\frac{(1+\alpha)(1-\cos\omega)}{\alpha^{2}}\right).\label{eq_ap2_2}
\end{equation}
The equation (\ref{eq_ap2_2}) defines a curve on the complex plane
on which the weak multipliers reside. For $\alpha>-1$ the curve resides
within the unity circle, which means that the weak spectrum is stable.
For $\alpha<-1$ the curve lies outside of the unity circle, and the
complete weak spectrum becomes unstable at once.

\section{Stability of multipartite solutions\label{sec:Stability-of-quasi-reappearant}}

A multipartite solution is a $P+1$-periodic solution $(\overline{T_{1}^{\ast},...,T_{P+1}^{\ast}})$
of the ISI map (\ref{eq:T-OIOU}). Equivalently, it corresponds to
a $P+1$-periodic point $\boldsymbol{T}^{\ast}=(T_{2}^{\ast},...,T_{P+1}^{\ast})$
of the $P$-dimensional return map $\boldsymbol{T}\mapsto\boldsymbol{T}^{{\rm new}}=R\left(\boldsymbol{T}\right)$
given by (\ref{eq:p-dim-map}). As such we study its stability, which
can be determined from the spectrum of the Jacobian of $R^{P+1}$
\begin{align}
{\cal J} & =D_{\boldsymbol{T}}\left[R^{P+1}\right](\boldsymbol{T}^{\ast})\nonumber \\
 & =D_{\boldsymbol{T}}R(R^{P}(\boldsymbol{T}^{\ast}))\cdot D_{\boldsymbol{T}}R(R^{P-1}(\boldsymbol{T}^{\ast}))\cdots D_{\boldsymbol{T}}R(\boldsymbol{T}^{\ast}).\label{eq:Jacobian-poincare1}
\end{align}
Since the nonlinear part of $R\left(\boldsymbol{T}\right)$ depends
exclusively on the sum $\sum T_{k}$, its Jacobian does so equally,
i.e., 
\[
D_{\boldsymbol{T}}R(\boldsymbol{T})=D\left(\alpha\right),
\]
where $\alpha=Z^{\prime}\left(\tau-\sum T_{k}\right)$ and $D(\alpha)$
is given by Eq.~(\ref{eq:Jacobian}). For a $P+1$-periodic point
$\boldsymbol{T}^{\ast}$, one application of the return map yields
\[
R\left(\boldsymbol{T}^{\ast}\right)=(T_{3}^{\ast},...,T_{P+1}^{\ast},T_{1}^{\ast}),
\]
and in general, after $k$ applications, we obtain
\[
R^{k}\left(\boldsymbol{T}^{\ast}\right)=(T_{\left[1+k\right]+1}^{\ast},...,T_{\left[P+k\right]+1}^{\ast}),
\]
with $\left[n\right]:=n\,\mathrm{mod}(P+1).$ Therefore, the factor
$D_{\boldsymbol{T}}R(R^{k}(\boldsymbol{T}^{\ast}))$, $k=0,...,P,$
of ${\cal J}$ is given by (\ref{eq:Jacobian}) with 
\begin{equation}
\alpha=\alpha_{k+1}=4Z^{\prime}\left(\tau-\sum_{j=1}^{P+1}T_{j}^{\ast}+T_{k+1}^{\ast}\right).\label{eq:alpha}
\end{equation}
The matrix ${\cal J}$ is therefore given as 
\begin{equation}
{\cal J}=D(\alpha_{P+1})\cdot D(\alpha_{P})\cdots D(\alpha_{1}).\label{eq:Jacobian-poincare2}
\end{equation}
Its spectrum can be obtained numerically. Due to (\ref{eq:alpha}),
for $T_{j}^{\ast}=T_{k}^{\ast},$ we have $\alpha_{j}=\alpha_{k}$.
Although we lack a proof, it seems that the spectrum of (\ref{eq:Jacobian-poincare2})
does not depend on the order of the factors $D(\alpha_{k})$, such
that the stability of solutions on all overlapping branches in Fig.~\ref{fig:elip}
coincides. More precisely, we conjecture that the characteristic polynomial
of (\ref{eq:Jacobian-poincare2}) has the form
\begin{align*}
\det\left({\cal J}-\lambda\right)= & \lambda^{P}-\left(s_{2}+...+s_{P+1}\right)\lambda^{P-1}-...\\
 & ...-\left(s_{P}+s_{P+1}\right)\lambda^{1}-s_{P+1},
\end{align*}
where $s_{j}=s_{j}(\alpha_{1},...,\alpha_{P+1})$ is the $j$-th symmetric
function. For instance, $s_{1}=\sum_{k=1}^{P+1}\alpha_{k}$, $s_{2}=\sum_{k=1,j>k}^{P+1}\alpha_{k}\alpha_{j}$,
and $s_{P+1}=\prod_{k=1}^{P+1}\alpha_{k}$.

\end{document}